\documentclass[aps,prl,twocolumn,groupedaddress,showpacs,superscriptaddress,amssymb,amsmath,noeprint,longbibliography]{revtex4-2}
\usepackage{graphicx}
\usepackage{dcolumn}
\usepackage{bm}
\usepackage{hyperref}
\usepackage{cleveref}
\hypersetup{
    colorlinks=true,
    linkcolor=blue,
    urlcolor=blue,
	citecolor=blue
}
\usepackage{comment}
\usepackage{color}
\usepackage[utf8]{inputenc}
\usepackage{graphicx}
\usepackage{tabularx}
\usepackage{xcolor}
\usepackage{amsmath}
\usepackage{dcolumn}
\usepackage{bm}
\usepackage{epsf}
\usepackage{braket}
\usepackage{tensor}
\usepackage{soul}
\usepackage{bbm}
\usepackage{tikz}
\usepackage[normalem]{ulem}

\begin{document}

\title{Accelerated relaxation and Mpemba-like effect  for operators in open quantum systems}

\author{Pitambar Bagui}
\email{pitambar.bagui@students.iiserpune.ac.in} 
\affiliation{Department of Physics, Indian Institute of Science Education and Research, Pune 411008, India}

\author{Arijit Chatterjee}
\email{arijitchattopadhyay01@gmail.com} 
\affiliation{Department of Physics and NMR Research Center, Indian Institute of Science Education and Research, Pune 411008, India}

\author{Bijay Kumar Agarwalla}
\email{bijay@iiserpune.ac.in}
\affiliation{Department of Physics, Indian Institute of Science Education and Research, Pune 411008, India}

\date{\today}

\begin{abstract}
Quantum Mpemba effect occurs when a quantum system, residing far away from the steady state, relaxes faster than a relatively nearer state. We look for the presence of this highly counterintuitive effect in the relaxation dynamics of the operators within the open quantum system setting. Since the operators evolve under a non-trace preserving map, the trace distance of an operator 
is not a monotonically decaying function of time, unlike its quantum state counterpart. Consequently, the trace distance can not serve as a reliable measure for detecting the Mpemba effect in operator dynamics. 
We circumvent this problem by defining a \textit{dressed} distance between operators that decays monotonically with time, enabling a generalized framework to explore the Mpemba-like effect for operators. Applying the formalism to various open quantum system settings, we find that, interestingly, in the single qubit case, only accelerated relaxation of operators is possible, while genuine Mpemba-like effects emerge in higher-dimensional systems such as qutrits and beyond. Furthermore, we demonstrate the existence of Mpemba-like effects in nonlocal, non-equilibrium operators, such as current, in a double-quantum-dot setup. Our results, besides offering fundamental insight about the occurrence of the Mpemba-like effect under non-trace preserving dynamics, open avenues for new experimental studies where quicker relaxation of observables could be of significant interest. 
\end{abstract}

\maketitle

{\it Introduction.--} The Mpemba effect refers to the bizarre observation \cite{E_B_Mpemba_1969,10.1119/1.1975687} that, under certain conditions, a hotter substance can cool down and relax faster than an initially cooler one when brought into contact with a cold reservoir.
Even after a series of theoretical and experimental investigations over many decades \cite{Greaney2011,Ahn2016,PhysRevLett.119.148001,Keller_2018,Hu2018,doi:10.1073/pnas.1819803116,chaddah2010overtakingapproachingequilibrium,PhysRevLett.129.138002,Holtzman2022,PhysRevLett.131.017101,PhysRevLett.132.117102,PhysRevLett.124.060602,walker2023mpembaeffecttermsmean,walker2023optimaltransportanomalousthermal,bera2023effectdynamicsanomalousthermal,PhysRevE.109.044149,PhysRevLett.134.107101,teza2025speedupsnonequilibriumthermalrelaxation,burridge2020observing,Vynnycky2010}, the precise conditions for the occurrence of this effect in classical systems continue to be debated \cite{Burridge2016,Bechhoefer2021}. However, the success in achieving a satisfactory understanding of the Mpemba effect \cite{doi:10.1073/pnas.1701264114,Kumar2020,doi:10.1073/pnas.2118484119} within the framework of Markovian dynamics has inspired investigations in quantum systems. When a quantum system, weakly coupled to a sufficiently large reservoir, is perturbed from equilibrium, it relaxes to the steady state following a Gorini–Kossakowski–Sudarshan–Lindblad (GKSL)  quantum master equation, which is known to be a completely positive trace-preserving (CPTP) map \cite{lidar2020lecturenotestheoryopen,breuer2002theory}. At a given time $t$ during the relaxation process, the state of the system $\rho(t)$ reads
\begin{equation}
\rho(t) = e^{\mathcal{L}t} \rho(0) = \rho_{\text{ss}} + \sum_{n=1}^{d^2-1} e^{\lambda_nt} a_n\,r_n, 
\label{eq:forward_map} 
\end{equation}
where $a_n=\text{Tr}[l_n\, \rho(0)]$ describes the overlap of the initial state with the $n$-th decay mode, while $l_n$ and $r_n$ being the respective left and right eigenoperators of the Liouvillian $\mathcal{L}$ and $d$ is the dimension of the Hilbert space. Considering a monotonically decaying distance function $\mathcal{D}(\rho(t),\rho_{\text{ss}})$ that quantifies how far the system $\rho(t)$ is from its long-time steady-state $\rho_{\text{ss}}$, one can formally define quantum Mpemba effect \cite{PhysRevLett.127.060401,PhysRevA.106.012207,bao2025acceleratingquantumrelaxationtemporary,PhysRevE.108.014130,PhysRevResearch.3.043108,PhysRevResearch.6.033330,PhysRevLett.133.136302,PhysRevLett.131.080402,PhysRevA.110.022213,PhysRevLett.133.140404,Ares2025,PhysRevA.110.042218,Longhi2025mpembaeffectsuper,PhysRevA.111.022215,zhao2025noiseinducedquantummpembaeffect,chang2024imaginarytimempembaeffectquantum,longhi2024bosonic,Kheirandish_2025,PhysRevB.111.125404,PhysRevLett.134.220402,Furtado_2025}. In essence, when initially $\mathcal{D}(\rho^{\text{far}},\rho_{\text{ss}}) > \mathcal{D}(\rho^{\text{near}},\rho_{\text{ss}})$, but the state $\rho^{\text{far}}$ reaches equilibrium faster than $\rho^{\text{near}}$, the system is said to exhibit quantum Mpemba effect.
Based on this formulation, the quantum Mpemba effect has recently been observed experimentally \cite{PhysRevLett.133.010402,Liang_2025} in the relaxation of ion traps \cite{Zhang2025,PhysRevLett.133.010403} and in the natural thermalization of nuclear spin systems \cite{chatterjee2025directexperimentalobservationquantum}.

Even though the occurrence of the Mpemba effect, or accelerated relaxation of quantum states, has been extensively investigated in recent years \cite{PhysRevLett.127.060401,PhysRevA.106.012207,bao2025acceleratingquantumrelaxationtemporary,PhysRevE.108.014130,PhysRevResearch.3.043108,PhysRevResearch.6.033330,PhysRevLett.133.136302,PhysRevLett.131.080402,PhysRevA.110.022213,PhysRevLett.133.140404,Ares2025,PhysRevA.110.042218,Longhi2025mpembaeffectsuper,PhysRevA.111.022215,zhao2025noiseinducedquantummpembaeffect,chang2024imaginarytimempembaeffectquantum,longhi2024bosonic,Kheirandish_2025,PhysRevB.111.125404,PhysRevLett.134.220402,Furtado_2025,PhysRevLett.134.220403}, studies focusing on dynamics of quantum operators remain largely unexplored. The existence of Mpemba-like effects in operator dynamics is of both operational and practical importance, as one is often interested in the accelerated relaxation of a particular observable \cite{PhysRevX.7.031027,PhysRevE.90.012121,Alpino_2025}, such as magnetization or energy current, rather than the relaxation of the full quantum state. Moreover, such investigations pose an intriguing fundamental challenge, since, unlike quantum states, the dynamical map governing operator evolution is generally non trace-preserving (nTP) \cite{breuer2002theory}. As a consequence, the trace distance measure $\mathcal{D}_{\text{tr}}(\rho,\rho_{\text{ss}})=\text{Tr} |\rho-\rho_{\text{ss}}|/2$ (where $|X|=\sqrt{X^{\dagger}X}$) \cite{nielsen2010quantum,Ares2025}, a standard metric to detect Mpemba effect for quantum states, becomes ill-suited for analyzing Mpemba effects for operators. This is because, under an nTP map, the trace distance can exhibit non-monotonic behavior, potentially leading to a false detection of Mpemba-like effects.

In this Letter, we show how this limitation can be circumvented. We introduce a \textit{dressed} distance that decays monotonically even under a nTP map, thereby serving as 
a suitable metric for investigating the Mpemba-like effect in operator dynamics of open quantum systems.  This weighted measure allows for a consistent formulation for Mpemba-like effects in the operator picture. We first apply this framework to a single-qubit setup and find that only accelerated relaxation of operators is possible. 
Motivated by this restriction, we extend our analysis to a qutrit system and demonstrate the emergence of genuine operator Mpemba-like effects. We further explore non-local operators, such as energy current, and observe both Mpemba-like behavior and accelerated relaxation in a double quantum dot (DQD) model with local Lindblad dissipators \cite{khandelwal2024emergentliouvillianexceptionalpoints}. 
Importantly, we show that the signatures of Mpemba-like effects, captured through the dressed distance, manifest directly in the time evolutions of corresponding operator expectation values.

\vspace{0.1cm}
{\it General theory of operator relaxation--} 
Given a GKSL quantum master equation, in the Heisenberg picture,  an operator $\mathcal{O}(t)$ evolves as $\mathcal{O}(t)=\exp(\mathcal{L}^{\dagger}t)\mathcal{O}(0)$, where the adjoint Liouvillian superoperator $\mathcal{L}^\dagger$ reads \cite{lidar2020lecturenotestheoryopen,breuer2002theory} 
\begin{equation}
    \mathcal{L}^{\dagger}[\mathcal{O}] = i[H, \mathcal{O}] + \sum_i \Gamma_i\left( L_i^{\dagger} \mathcal{O} L_i - \frac{1}{2} \{ L_i^\dagger L_i, \mathcal{O} \} \right).
    \label{eqn_adlind}
\end{equation}
Here $H$ is the Hamiltonian of the system of interest, $L_i, L_i^{\dagger}$ are the jump operators corresponding to the $i$-th channel, and $\Gamma_i$ is the associated rate.
Let us arrange the real part of the eigenvalues $\lambda_i$ of $\mathcal{L}^{\dagger}$ in descending order, i.e., $\{\lambda_0=0 > \text{Re}(\lambda_1) \geq \text{Re}(\lambda_2) ...\}$. We assume that the eigenspace of the generator of the quantum state $\mathcal{L}$ corresponding to the zero eigenvalue is non-degenerate, and denote the corresponding right and left eigenoperators by $r_0$ and $l_0$, respectively. 
The non-degeneracy of the eigenvalue $\lambda_0$ ensures an unique steady state $\rho_{\rm ss}=r_0$, and thus  an operator  $\mathcal{O}(t)$ converges to $\mathcal{O}_{\rm ss}=\mathrm{Tr}[r_0\mathcal{O}(0)] \, l_0$ as $t\rightarrow \infty$. Moreover, as the left eigenoperator $l_0$ is an identity operator $\mathbb{I}$, the steady-state operator is always proportional to $\mathbb{I}$. 
Following the spectral decomposition of $ \mathcal{L}^\dagger$,  
the time-evolved operator $\mathcal{O}(t)$ can be expressed as
\begin{equation}
    \mathcal{O}(t) = \mathcal{O}_{\mathrm{ss}} 
    + \sum_{i=1}^{d^2 - 1} c_i\, e^{\lambda_i t} \, l_i,
    \label{eq:operator-decomp}
\end{equation}
where the coefficients  
\(
c_i = \mathrm{Tr} \big[\mathcal{O}(0)\,r_i \big]
\)  
are the overlaps of the initial observable $\mathcal{O}(0)$ with the right eigenoperators $r_i$ of $\mathcal{L}$.  

We now focus on investigating how quickly the operator $\mathcal{O}(0)$ relaxes to $\mathcal{O}_{\mathrm{ss}}$. For that purpose, we define a \emph{dressed distance} between them as
\begin{equation}
    \mathcal{D}_{\rm dd}(\mathcal{O}(t), \mathcal{O}_{\mathrm{ss}}) = \mathrm{Tr} \,\big| \mathcal{O}(t) - \mathcal{O}_{\mathrm{ss}} \big|_{\rho_{\rm{ss}}},\label{dressed_norm}
\end{equation}
where for an arbitrary Hermitian operator $X$, $|X|_{\rho_{\rm{ss}}}$ is defined as
\begin{equation}
|X|_{\rho_{\rm{ss}}} = \sqrt{\big(\sqrt{\rho_{\rm{ss}}}\, X \, \sqrt{\rho_{\rm{ss}}}\big)^{\dagger}\big(\sqrt{\rho_{\rm{ss}}}\,X\,\sqrt{\rho_{\rm{ss}}}}\big).
\label{eq:sqrt_mat}
\end{equation}
Interestingly, it can be shown that for any two time instants $t_2>t_1$, the dressed distance $\mathcal{D}_{\mathrm{dd}}$ decays monotonically with time, i.e.,
\begin{equation}
\mathcal{D}_{\mathrm{dd}}(\mathcal{O}(t_2), \mathcal{O}_{\mathrm{ss}}) \leq \mathcal{D}_{\rm dd}(\mathcal{O}(t_1), \mathcal{O}_{\mathrm{ss}}).
\label{eq:mon_decay}
\end{equation}
[we refer to \cite{supp} for the details of the proof]. It is important to emphasize that, if the standard trace distance is used instead of the dressed distance as a metric for operators, a monotonic decay, as expressed in Eq.~\eqref{eq:mon_decay}, is not generally guaranteed \cite{supp}. This unique property of the dressed distance is therefore crucial for reliably identifying Mpemba-like effects in operator dynamics of open quantum systems.

\vspace{0.1cm}
{\it Accelerated relaxation and Mpemba-like effect for operators.--}
The time an operator takes to reach its steady-state crucially depends on its overlaps $\{c_i\}$ with the decay modes, as defined in Eq.~\eqref{eq:operator-decomp}.
At long times, the dynamics is dominated by the slowest decay mode, characterized by the eigenvalue $\lambda_1$. If the initial operator exhibits zero overlap with this mode, its contribution vanishes, resulting in accelerated relaxation toward the steady state.
In such cases, the relaxation timescale is governed by the next-slowest decay mode, characterized by eigenvalue $\lambda_2$, (assuming $\lambda_1$ as real). However, if the slowest mode is complex, then $\lambda_1$ and $\lambda_2$ form a complex conjugate pair with equal real parts, resulting in the same decay rate. To observe accelerated relaxation, in this case, the operator must have zero overlap with both $\lambda_1$ and $\lambda_2$ decay modes. A practical approach to bypass the contribution of the slowest decay mode is to apply a suitable unitary transformation to the operator such that 
\begin{equation}
\label{zero-overlap}
    \mathrm{Tr}[\widetilde{\mathcal{O}}(0)\, r_1] = \mathrm{Tr} [U^\dagger \, \mathcal{O}(0)\, U\, r_1] = 0,
\end{equation}
which will lead to an accelerated relaxation. Here we assume that $\lambda_1$ is real. For the case when $\lambda_1$ and $\lambda_2$ are complex conjugate pairs, one need to satisfy $\mathrm{Tr}(U^\dagger \mathcal{O}(0) U\, r_1) = \mathrm{Tr}(U^\dagger \mathcal{O}(0) U\, r_2) = 0$ to observe accelerated relaxation. Moreover, along with satisfying the relation in Eq.~\eqref{zero-overlap}, if the action of the unitary $U$ also pushes the operator $\mathcal{O}$ further away relative to $\mathcal{O}_{\text{ss}}$, i.e., in terms of dressed distance [Eq.~\eqref{dressed_norm}] if 
$\mathcal{D}_{\rm dd}(\widetilde{\mathcal{O}}(0), \widetilde{\mathcal{O}}_{\mathrm{ss}})> 
\mathcal{D}_{\rm dd}(\mathcal{O}(0), \mathcal{O}_{\mathrm{ss}})$, then a Mpemba-like effect can be observed for the operator. Note that, in general, such a unitary operator $U$ may not always exist or difficult to construct. In such a case, one may still find a unitary that reduces the overlap of $\mathcal{O}(0)$ with the slowest decay mode, which leads to a weak Mpemba-like effect or partial acceleration for the operator's relaxation.

At this juncture, it is important to note that, owing to the trace-preserving property of the Liouvillian $\mathcal{L}$ governing the evolution of a quantum state, a unique steady state $\rho_{\text{ss}}$, independent of the initial condition always exists (assuming non-degenerate $\lambda_0$).  In contrast, under the action of $\mathcal{L}^{\dagger}$, an operator $\mathcal{O}$ relaxes  to a steady-state  $\mathcal{O}_{\text{ss}}=\text{Tr}[r_0\mathcal{O}(0)]\mathbb{I}$, which depends explicitly on the initial operator $\mathcal{O}(0)$. 
Consequently, for quantum states, the steady-state $\rho_{\rm ss}$ remains invariant under any unitary transformation of the initial state. However, for operator this invariance no longer holds; before and after a unitary transformation the respective steady-states are $\mathcal{O}_{\text{ss}}=\text{tr}[r_0\mathcal{O}(0)]\mathbb{I}$, and $\widetilde{\mathcal{O}}_{\text{ss}}=\text{tr}[r_0 \, \widetilde{\mathcal{O}}(0)]\mathbb{I}$, respectively. It is in this sense that we refer to the anomalous relaxation of operators as a \textit{Mpemba-like effect}, reserving the term Mpemba effect for quantum state relaxation.

If, however, we require the steady-state operator to remain unchanged before and after the unitary transformation i.e., $\widetilde{\mathcal{O}}_{\rm ss}= \mathrm{Tr}\!\left[\rho_{\mathrm{ss}}\, U^{\dagger}\mathcal{O}(0)U\right] \mathbb{I}= \mathcal{O}_{\rm ss}$, a sufficient condition for this to hold is
\begin{equation}
    \Big[U, \rho_{\mathrm{ss}}\Big] = 0.  \label{eq:same_sted_ste}
\end{equation}
Interestingly, under this condition, the 
initial dressed distance before and after the unitary also remains identical, $\mathcal{D}_{ \mathrm{dd}}(\mathcal{O}(0),\mathcal{O}_{\rm{ss}})=\mathcal{D}_{\mathrm{dd}}(\widetilde{\mathcal{O}}(0),\mathcal{O}_{\rm{ss}})$ [we refer to \cite{supp} for the details of the proof]. As a consequence, in such scenarios, only accelerated relaxation is possible, and {\it no} Mpemba-like effect can occur. A simple yet insightful example arieses when the Liouvillian $\mathcal{L}$ corresponds to a unital map, for which $\rho_{\mathrm{ss}}=\mathbb{I}/d$. In this case Eq.~\eqref{eq:same_sted_ste} is trivially satisfied. Nevertheless, certain carefully chosen unitary transformation can still lead to accelerated relaxation. In other words for a unital map, only accelerated relaxation for operators is possible and no Mpemba effect can occur.

In what follows, we illustrate the above results in the single-qubit setting, where, interestingly, only accelerated relaxation is possible. We then extend our analysis to a qutrit system, where Mpemba-like effects emerge, and further to a minimally extended setup -- a boundary-driven double quantum dot that supports a non-equilibrium steady state and demonstrates Mpemba-like effects through the dynamics of non-equilibrium current operator. 

\begin{figure}
    \centering
    \includegraphics[trim=0cm 0cm 0cm 0cm, clip=true,width=8.5cm]{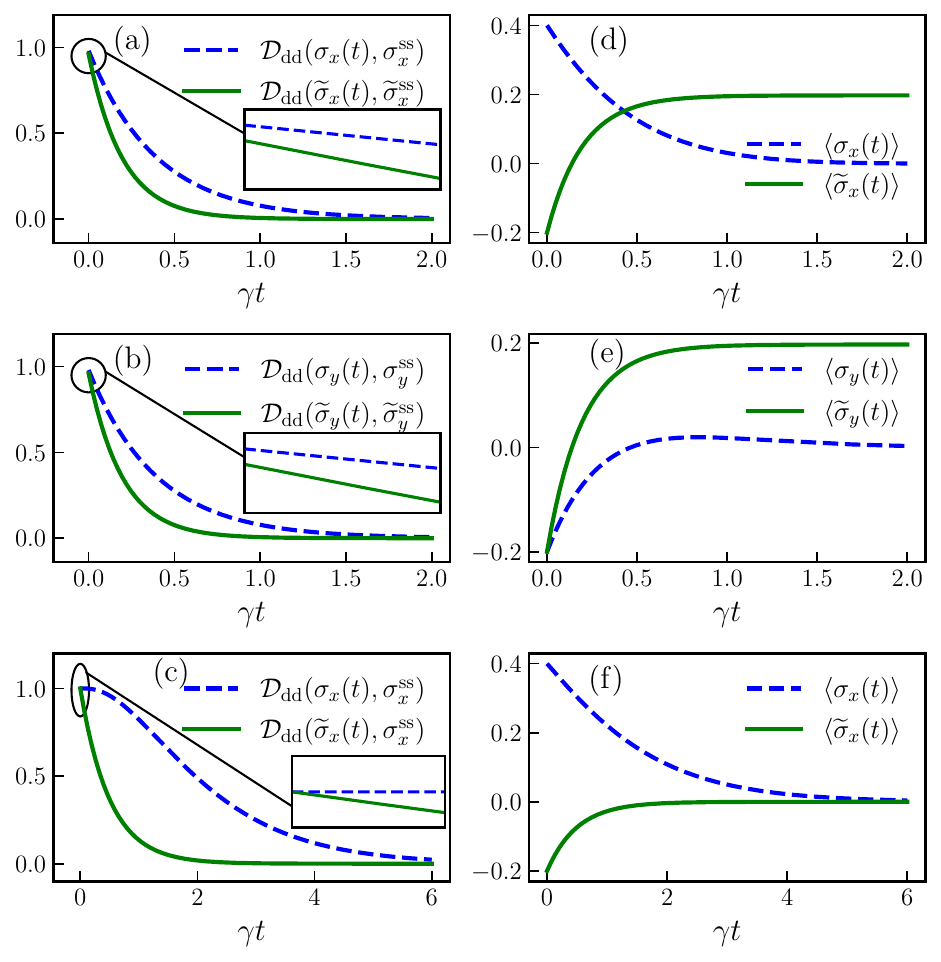}
    \caption{(a-c) Monotonic decay of the dressed distance, defined in Eq.~\eqref{dressed_norm}, for single-qubit Pauli operators before (dashed line) and after (solid line) applying a unitary transformation. For (a) and (b) we choose the dissipator in Eq.~\eqref{eqn_adlind} as $\sigma_{+}$ and $\sigma_{-}$ with rates satisfying the detailed balance condition. For (c) the dissipator is chosen as $\sigma_x$. In all cases, only accelerated relaxation is observed, as the initial dressed distance after unitary transformation is always less or equal (such as for unital map) than before applying unitary transformation, as shown clearly in the inset (a)-(c). (d-f) The signature of the respective operator relaxation is reflected in the time evolutions of their expectation values computed for a random initial state $\rho_0$. However, this signature is also reflected for an arbitrary chosen initial state.  For (d)-(e), the operators before and after unitary protocol, relaxes to different steady-states, whereas for the unital map with $\sigma_x$ dissipator in (f), the operators before and after unitary relaxes to the same steady state.    The simulations are done using parameters $\gamma = 1$, $\omega_0 = 1$, $k_BT = 2$ for dissipators $\sigma_{\pm}$ and $\gamma=1$, $\omega_0=1$ for dissipator $\sigma_{x}$.}
\label{fig:single_qubit_trace_dist_acceleration}
\end{figure}

\vspace{0.1cm}
{\it Example 1: Single qubit.--}
We first consider a single qubit system described by the Hamiltonian: $H = \omega_0 \, \sigma_z/2$,
where $\sigma_z=\Ket{1}\Bra{1}-\Ket{0}\Bra{0}$ is the Pauli-Z operator and $\omega_0$ is the energy gap between this two level. The GKSL dynamics of any operator $\mathcal{O}$ follows Eq.~\eqref{eqn_adlind}. We choose the jump operators as $\sigma_+=\Ket{1}\Bra{0}$ and $\sigma_-=\sigma_{+}^{\dagger}$ with corresponding rates $\gamma n_b$ and $\gamma(1 + n_b)$, respectively. Here $n_b=1/(\exp(\hbar\omega_0/k_BT)-1)$ is the Bose-Einstein occupancy factor at bath temperature $T$, and $k_B$ is the Boltzmann constant.  In this case,  the eigenvalues $\lambda_1$ and $\lambda_2$ of the adjoint Liouvillian form a complex-conjugate pair ($\lambda_2=\lambda_1^{\star}$) corresponding to the slowest decay mode (with equal real parts). The associated right eigenmatrices $r_1$ and $r_2$ are complex conjugates of each other and are purely off-diagonal, a characteristic feature of the Davies map structure \cite{davies1979generators, PhysRevLett.133.140404}. Therefore, to 
suppress the contribution of these slowest decaying modes and thereby achieve accelerated relaxation,
it is sufficient to make the operator $\mathcal{O}$  diagonal through an appropriate unitary transformation $U$, such that $\mathrm{Tr}(U^\dagger \mathcal{O}(0) U\, r_1) = \mathrm{Tr}(U^\dagger \mathcal{O}(0) U\, r_2) = 0$. 
In Fig.~\ref{fig:single_qubit_trace_dist_acceleration}~(a-b), we plot the time evolution of the dressed distance  by choosing (a) $\mathcal{O}=\sigma_x$, (b) $\mathcal{O}=\sigma_y$, and observe a clear speed up in relaxation after applying unitary transformation (solid line), in both cases. (d)-(e) shows the corresponding accelerated relaxation behaviour in the expectation values of the operators. Interestingly, for the single qubit setup, for an initial operator $\mathcal{O} = b_0 \, \mathbb{I} + \vec{b} \cdot \vec{\sigma}$, 
where $\vec{b} \in \mathbb{R}^3$, the initial dressed distance is always larger than the one involving the unitarily transformed diagonal operator $\mathcal{\widetilde{O}} = b_0 \, \mathbb{I} + b \, \sigma_z$, where $b=\sqrt{b_x^2 + b_y^2 + b_z^2}$.
In other words, $\mathcal{D}_{\rm dd}(\widetilde{\mathcal{O}}(t), \widetilde{\mathcal{O}}_{\mathrm{ss}}) < \mathcal{D}_{\rm dd}(\mathcal{O}(t), \mathcal{O}_{\mathrm{ss}})$ implying that \textit{no} Mpemba-like effect occurs in the single qubit case [we refer to \cite{supp} for the details of the proof]. In this setup, with the choice of $\sigma_{\pm}$ dissipators, the operators relax to different steady state values before and after the unitary transformation, as shown in (d)-(e). One can however achieve accelerated relaxation with the same steady-state. This is what we show in Fig.~\ref{fig:single_qubit_trace_dist_acceleration}~(c) where we consider $\sigma_x$ as the only jump operator which generates an unital map with steady state $\rho_{ss}=\mathbb{I}/2$. As discussed earlier, for this case, the initial dressed distance before and after unitary also becomes the same and Eq.~\eqref{eq:same_sted_ste} gets satisfied for a chosen $U$, which can provide accelerated relaxation. The corresponding expectation values of the operators also shows quick relaxation and with the same steady-state value. As the Mpemba-like effect is absent in this model, we now extend our analysis to a qutrit setup.

\begin{figure}
\centering
\includegraphics[trim=0.2cm 0.25cm 0.2cm 0.25cm, clip=true,width=8cm]{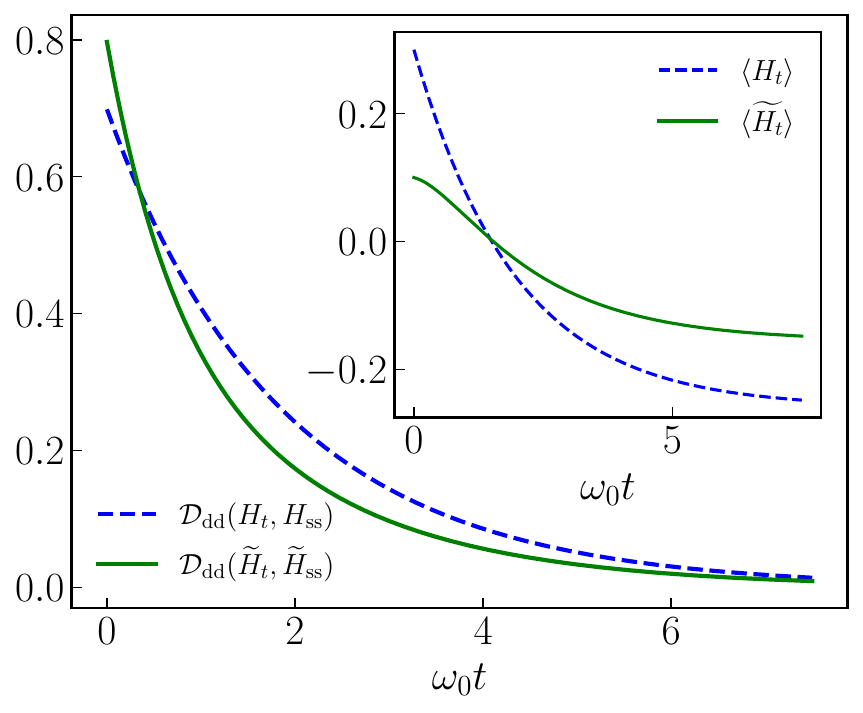}
\caption{Mpemba-like effect in the time dynamics of dressed distance for Hamiltonian operator for a qutrit setup, where the unitary transformed Hamiltonian i.e., $\widetilde{H}=U^{\dagger}HU$ relaxes much faster to $\widetilde{H}_{\text{ss}}$ than the corresponding relaxation of $H$ to $H_{\text{ss}}$, even though initially $\mathcal{D}_{\rm dd}(\widetilde{H}(0),\widetilde{H}_{\text{ss}})>\mathcal{D}_{\rm dd}(H(0), H_{\text{ss}})$. The corresponding time evolution of the expectation values (inset) of the respective operators also carries the signature of the Mpemba-like effect.  Here, the average is computed using a diagonal initial state $\rho_0={\mathrm{diag}} (0.5,0.3,0.2)$. 
The parameters for the setup are chosen as $\omega_0=1$, $\gamma=0.2$, $k_B T=2.5$. }
\label{threelevel_mpemba}
\end{figure}

\vspace{0.1cm}
\noindent {\it Example 2: Single qutrit .--}
We next consider a spin-$1$ system given by the  Hamiltonian
$H = \omega_0 S_z$. The eigen-states are given as $H\Ket{i}= \omega_0\Ket{i}$, for $i\in\{1,0,-1\}$. 
The dissipators that we consider here are  $L_{+}^{x(y)}=\Ket{0(1)}\Bra{-1(0)}$ and $L_{-}^{x(y)}=\Ket{-1(0)}\Bra{0(1)}$ with respective decay rates $\Gamma_{+}^{x(y)}=\gamma \, n_b$ and $\Gamma_{-}^{x(y)}=\gamma\,(1+ n_b)$. Since such a scenario corresponds to the Davies map
\cite{davies1979generators,PhysRevLett.133.140404}, the unitary and dissipative parts commute, and also the population and coherence terms evolve independently. In particular, the right and left eigenmatrices of $\mathcal{L}$ corresponding to the complex eigenvalues 
$\{\lambda_1,\lambda_2=\lambda_1^{\star},\lambda_4,\lambda_5=\lambda_4^{\star},\lambda_6,\lambda_7=\lambda_6^{\star}\} \subset \mathbb{C}$ are purely off-diagonal and hence causes evolutions of coherence terms while populations evolve via the diagonal left and right eigenmatrices corresponding to real eigenvalues $\{\lambda_0,\lambda_3,\lambda_8 \}\subset \mathbb{R}$.

In this case, we study the relaxation of the Hamiltonian operator $H$. Since $H$ is diagonal, it overlaps with the eigenmodes corresponding to the eigenvalues $\lambda_0,\lambda_3$ and $\lambda_8$. Applying a unitary operator to $H$ that swaps its first and second diagonal elements reduces the overlap with $r_3$ while increasing the overlap with $r_8$. As can be seen in Fig.~\ref{threelevel_mpemba}, even though the initial transformed operator's dressed distance is much higher than that of $H$, it relaxes much faster, and a crossing is observed. Thus, Fig.~\ref{threelevel_mpemba} displays a Mpemba-like effect in the relaxation dynamics of the operator for a qutrit setup.

\vspace{0.1cm}
{\it Example 3: Double quantum dot setup.--} So far, our analysis has been restricted to local spin systems. However, the Mpemba-like effect can also emerge in extended setups that are driven out-of-equilibrium. To illustrate this, we next consider a minimal non-equilibrium extended system-- a boundary-driven double quantum dot (DQD). The Hamiltonian of the DQD is given as,
\begin{equation}
    {H}_{\mathrm{DQD}}
    = \epsilon_{\mathrm{d}1}{n}_1
    + \epsilon_{\mathrm{d}2}{n}_2
    + g \left( {c}^{\dagger}_1 {c}_2 + {c}^{\dagger}_2 {c}_1 \right),
    \label{DQD-Ham}
\end{equation}
where $n_i=c_i^{\dagger} c_i$ is the occupation number operator of the $i$-th site with $c_i$ ($c^{\dagger}_i$) being annihilation (creation) operator of the $i$-th dot, satisfying the fermionic anti-commutation relation $\{c^{\dagger}_i,c_j\}=\delta_{ij}$. The parameters $\epsilon_{\mathrm{d}1}$ ($\epsilon_{\mathrm{d}2}$) represent the on-site energies of the left (right) dot, while $g$ denotes the interdot coupling strength. 
Each dot is coupled to its own fermionic reservoir with respective Hamiltonian $H^{(j)}_{R} = \sum_r \epsilon_{rj}\, c^{\dagger}_{rj}c_{rj}$, and is characterized by a chemical potential $\mu_j$ and a temperature $T_j$, for $j=1,2$.  
The DQD--reservoir coupling is given by $H^{(j)}_{\mathrm{DQD}R}= \sum_r t_{rj}\left( c^{\dagger}_{rj}c_{j} + c^{\dagger}_{j}c_{rj} \right)$,
where $c_{rj}$ corresponds to the $r$th mode of reservoir $j$, coupled to the $j$th quantum dot via the tunneling amplitude $t_{rj}$. We consider the Jordan-Wigner transformation \cite{schaller2014open} to map this system to spins such that the Hamiltonian in Eq.~\eqref{DQD-Ham} becomes $
    H_{\mathrm{DQD}}
    = \epsilon_{\mathrm{d}1}\,\sigma^{(1)}_+ \sigma^{(1)}_- \otimes \mathbb{I}_2
    + \epsilon_{\mathrm{d}2}\,\mathbb{I}_2 \otimes \sigma^{(2)}_+ \sigma^{(2)}_- 
     + g\left(\sigma^{(1)}_+ \otimes \sigma^{(2)}_- 
    + \sigma^{(1)}_- \otimes \sigma^{(2)}_+\right)
$.  At weak values of $g$, and under the Born--Markov and secular approximations,  
the time evolution of an operator is governed by Eq.~\eqref{eqn_adlind}, with four jump operators $L_{\pm}^{1(2)}=\sigma_{\pm}^{1(2)}$, where $A^{k}=A \otimes \mathbbm{1}$ or $\mathbbm{1} \otimes A$ for $k=1$ or $2$, respectively. The corresponding decay rates read $\Gamma_{+}^{1(2)}=\gamma_{1(2)}f(\epsilon_{d1(2)})$, and $\Gamma_{-}^{1(2)}=\gamma_{1(2)}[1-f(\epsilon_{d1(2)})]$, where $f_j(\epsilon) = 1/[e^{\beta_j(\epsilon-\mu_j)}+1]$ is  the Fermi function ensuring the detailed-balance relation $\Gamma^{j}_+/\Gamma^{j}_- = \exp[{-\beta_j(\epsilon_{dj}-\mu_j)]}
$, with $\beta_j=1/(k_BT_j)$ and $\mu_j$ are the inverse temperature and chemical potential of the $j$'th bath, respectively. 
\begin{figure}
    \centering
\includegraphics[trim=0cm 0.8cm 1cm 1.8cm, clip=true,width=7.5cm]{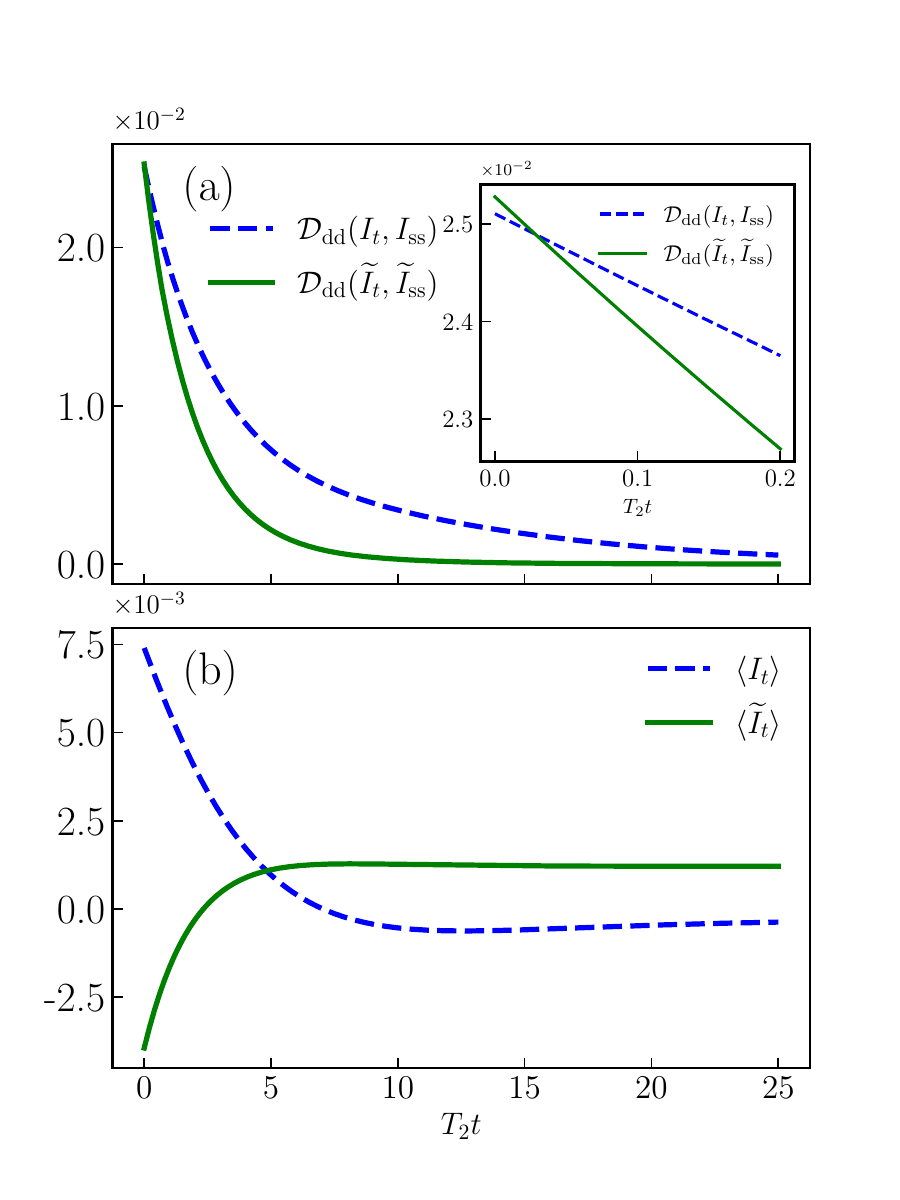}
    \caption{(a) Occurrence of the Mpemba-like effect in the relaxation dynamics of the current operator $I$, and its unitarily transformed counterpart $\widetilde{I}$ in the DQD model. Inset: enlarged view of the short-time dynamics highlighting the initial difference between the dressed distance. (b) Corresponding time evolution of the expectation values of $\langle I \rangle$ and $\langle \widetilde{I} \rangle$, computed for a random initial state $\rho_0$, exhibiting distinct slow and fast relaxation dynamics. The simulations are done with parameter values $\epsilon_{\mathrm{d}1}=0.2$, $\epsilon_{\mathrm{d}2}=0.1$, $g=0.05$, $k_BT_1=15$, $\mu_1=0$, $k_BT_2=1$, $\mu_2=0$, $\gamma_1=0.1$, $\gamma_2=0.5$}
    \label{DQD_mpemba_relax}
\end{figure}
\begin{figure}
    \centering
\includegraphics[trim=0cm 0cm 0cm 1.8cm, clip=true,width=7.5cm]{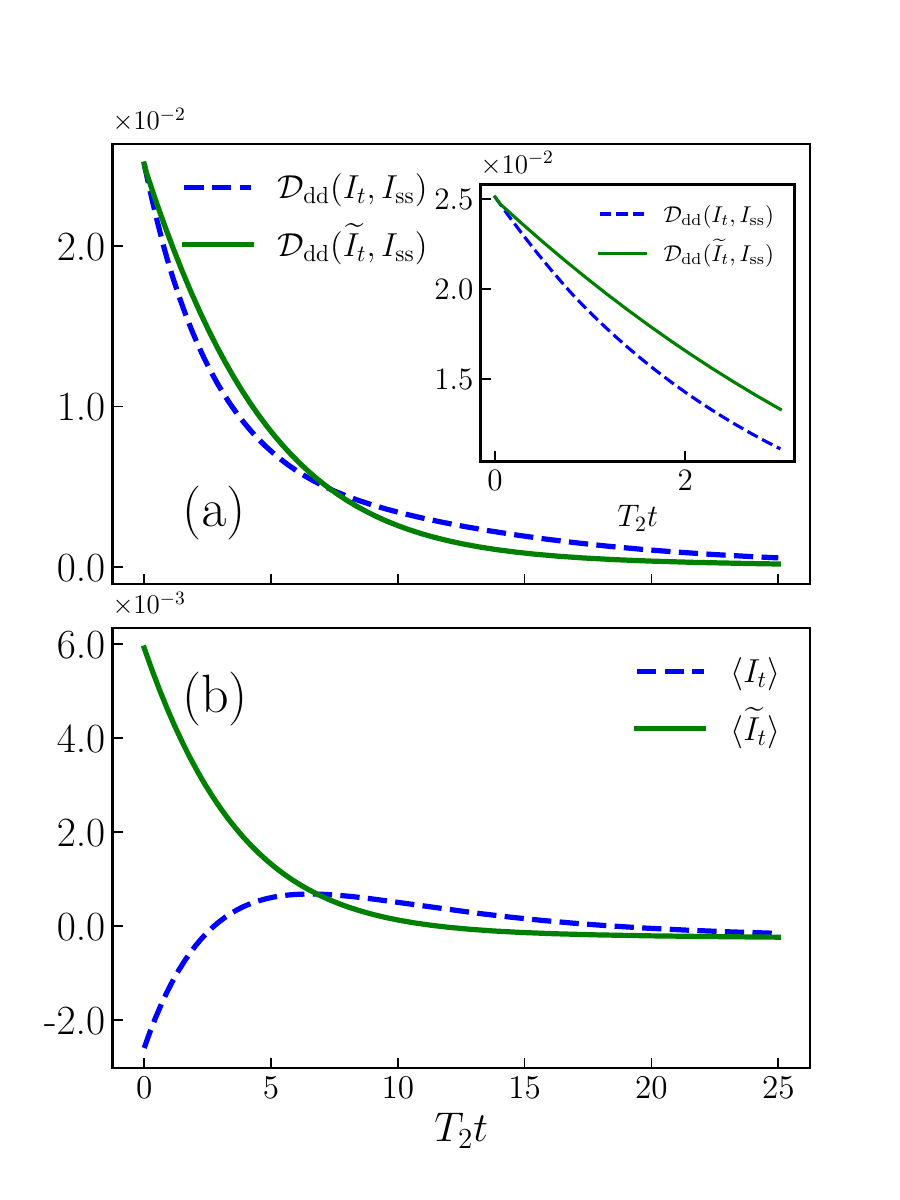}
    \caption{(a) Accelerated relaxation of the transformed operator $\widetilde{I}$ following a unitary transformation on the original current operator ${I}$, with both converging to the same steady-state operator at long times. (b) Corresponding operator dynamics reflected in the time evolutions of the respective expectation values, computed for a random initial state $\rho_0$.
    All simulations are done with parameter values $\epsilon_{\mathrm{d}1}=0.2$, $\epsilon_{\mathrm{d}2}=0.1$, $g=0.05$, $k_BT_1=1.5$, $\mu_1=0$, $k_BT_2=1$, $\mu_2=0$, $\gamma_1=0.1$, $\gamma_2=0.5$.}
    \label{DQD_accle_relax}
\end{figure}

For such a boundary-driven non-equilibrium setup, we focus on the energy current operator defined as 
$
I=ig\Big(\sigma^{(1)}_- \otimes\sigma^{(2)}_+-\sigma^{(1)}_+ \otimes \sigma^{(2)}_-\Big)
$,
and investigate the possible emergence of the Mpemba-like effect. The time evolution of $I$ under Eq.~\eqref{eqn_adlind} is given by
\begin{equation}
    I_{t}=\mathrm{Tr}(I\rho_{\mathrm{ss}})\,\mathbb{I} +\sum_{i=1}^{15}\mathrm{Tr}(r_i I)e^{\lambda_i t}l_i.
    \label{eq:curr_DQD}
\end{equation}
In this setup, the slowest decay mode corresponds to the eigenvalue $\lambda_3$, since the operator $I$ does not overlap with the modes $\lambda_{1}$ and $\lambda_2$. We apply a unitary transformation $U$ to $I$, obtaining the operator $\widetilde{I}$ with diagonal entries $\{-g,g,0,0\}$ which substantially reduces the overlap $\mathrm{Tr}(r_3 \tilde{I})$. As shown in Fig.~\ref{DQD_mpemba_relax}(a), although $\widetilde{I}$, starts farther from its steady-state, it relaxes significantly faster than the original operator $I$, thereby demonstrating the Mpemba-like effect. Fig.~\ref{DQD_mpemba_relax}~(b) further illustrates this behaviour in terms of the expectation values computed for a random initial state $\rho_0$, clearly showing that $\langle \widetilde{I}\rangle$ relaxes to its steady-state value more rapidly than $\langle I\rangle$. 
 
However, if the unitary $U$ applied to the operator $I$ satisfies Eq.~\eqref{eq:same_sted_ste}, both $I$ and the $\widetilde{I}$ converge to the same steady-state operator $I_{\text{ss}}$ at long times. In this case, as discussed earlier, the initial dressed distance of $I$ and $\widetilde{I}$ with respect to $I_{\text{ss}}$ remains the same. However $\widetilde{I}$ relaxes to $I_{\text{ss}}$ much faster than $I$, confirming accelerated relaxation, as shown in Fig.~\ref{DQD_accle_relax}~(a).
 The corresponding expectation values $\langle I\rangle$ and $\langle \widetilde{I} \rangle$, computed for a random initial state $\rho_0$, clearly captures this contrast between slow and fast relaxation, as shown in Fig~\ref{DQD_accle_relax}~(b). 

\emph{Summary and Outlook --} Quantum Mpemba effect, being a topic of utmost interest to researchers, has been widely studied in recent years, both for its fundamental importance and potential application in reducing the lag time between two consecutive runs of quantum devices via fast thermalization. Since any quantum phenomenon is primarily perceived via measuring the observables in experiments, we focus on studying the Mpemba effect in the relaxation dynamics of quantum operators themselves. As the map that evolves operators is not trace-preserving, usual distance functions used to detect state Mpemba do not remain monotonically decaying when used in operator dynamics. We surpass this problem by constructing a dressed norm that decays monotonically even under a nTP map, and use it to investigate the existence of quantum Mpemba-like effects in operator dynamics in qubits, qutrits, and a non-local double quantum dot model. Our approach can be suitably adapted for more complex setups to observe Mpemba-like effects or accelerated relaxation. Apart from the foundational impact on establishing the Mpemba-like effects in general operators under nTP dynamics, our work may find its application in various experimental scenarios where quicker relaxation of some particular observable  is required.

\emph{Acknowledgement.--} PB acknowledges Triyasa Kunti, Katha Ganguly, and Yash Pai for valuable discussions and the University  Grants Commission (UGC), Government of India, for the  research fellowship (Ref No.- 231620073714). AC acknowledges T.S. Mahesh for valuable discussions and Patankar Fellowship for financial support. BKA acknowledges CRG Grant No. CRG/2023/003377 from Science and Engineering Research Board (SERB), Government of India.  BKA acknowledges the hospitality of the International Centre of Theoretical Sciences (ICTS), Bangalore, India and International Centre of Theoretical Physics (ICTP), Italy, under the associateship program.

\bibliography{bibliography}

\onecolumngrid
\newpage
\begin{center}
\textbf{\large Supplemental Material for \\
``Accelerated relaxation and Mpemba-like effect  for operators in open quantum systems''}
\end{center}
\setcounter{section}{0}
\setcounter{figure}{0}
\setcounter{table}{0}
\setcounter{equation}{0}
\renewcommand{\theequation}{S\arabic{equation}}
\renewcommand{\thefigure}{S\arabic{figure}}
\renewcommand{\thetable}{S\arabic{table}}

\section*{A: Proof for monotonic decay of dressed distance under a non-trace preserving (n-TP) Markovian map}
In this section, we present the proof for the monotonic decay of dressed distance $\mathcal{D}_{\rm dd}(\mathcal{O}(t), \mathcal{O}_{\mathrm{ss}})$, defined in Eq.~\eqref{dressed_norm} of the main text, under a non-trace preserving (n-TP) Markovian map. We consider a quantum system with a Hilbert Space $\mathcal{H}$ attached to it. Let $\mathcal{B}(\mathcal{H})$ be the space of all bounded self-adjoint trace class operators on $\mathcal{H}$, i.e., $\mathcal{B}(\mathcal{H}) := \{\mathcal{O}:\mathcal{H}\rightarrow \mathcal{H} | ~\mathcal{O}=\mathcal{O}^{\dagger}, ~\rm{tr}(\mathcal{O}) < \infty, ~\rm{and \,\, the \,\, norm }~|\!|\mathcal{O}|\!| < \infty  \}$. We  assume that the system is undergoing a Markovian open system dynamics. In the Heisenberg picture, the evolution of any arbitrary operator $\mathcal{O} \in \mathcal{B}(\mathcal{H})$ will by the GKLS master equation, given in Eq.~\eqref{eqn_adlind} of the main text. The formal solution of the equation reads $\mathcal{O}(t)=\Phi_t \,\mathcal{O}(0) = \exp(\mathcal{L}^{\dagger}t)\,\mathcal{O}(0)$. Note that the map $\Phi_t=\exp(\mathcal{L}^{\dagger}t)$ is a complete-positive but a non-trace preserving (n-TP) map. Here $\mathcal{L}^{\dagger}$ is the adjoint Liouvillian. We assume that the eigenspace of generator $\mathcal{L}$ corresponding to the zero eigenvalue is non-degenerate, and the respective right and left eigen-matrices of $\mathcal{L}$ are $r_0$ and $l_0$, respectively. Note that $r_0$ is in fact steady-state density matrix $\rho_{\rm ss}$ and $l_0$ is an identity matrix. We define $\mathcal{O}_{\rm{ss}}:=\mathrm{Tr}[r_0 \, \mathcal{O}(0)]\,l_0$ as the operator to which $\mathcal{O}$ converges at long times. To study how quickly the operator $\mathcal{O}(0)$ relaxes to $\mathcal{O}_{\rm ss}$, we define a dressed distance between them as, 
\begin{align}
\!\mathcal{D}_{\mathrm{dd}}(\mathcal{O}(t),\mathcal{O}_{\rm{ss}}) &\!=\! |\!| \mathcal{O}(t),\mathcal{O}_{\rm{ss}} |\!|_{\rho_{\rm{ss}}} \!=\! |\!| \sqrt{\rho_{\rm{ss}}} \, \big(\mathcal{O}(t)\!-\!\mathcal{O}_{\rm{ss}}\big)\sqrt{\rho_{\rm{ss}}} |\!|,\nonumber\\
&=\mathrm{Tr}\big|\sqrt{\rho_{\rm{ss}}} \, (\mathcal{O}(t) - \mathcal{O}_{\mathrm{ss}}) \, \sqrt{\rho_{\rm{ss}}}\big|,
\label{eq:EM1}
\end{align}
 where $|\!| \,.\, |\!|_{\rho_{\rm ss}}$, as defined in the first line, denotes norm of an operator dressed by the steady-state density matrix corresponding to the Liouvillian ${\cal L}$, and $|X|=\sqrt{ X^{\dagger}X}$. For any Hermitian Operator $\mathcal{O}$ evolving under the map $\Phi_t=\exp(\mathcal{L}^{\dagger}t)$, we need to prove that $\mathcal{D}_{\mathrm{dd}}(\mathcal{O}(t_2),\mathcal{O}_{\rm{ss}}) \leq \mathcal{D}_{\rm dd}(\mathcal{O}(t_1),\mathcal{O}_{\rm{ss}}) $, where $t_2>t_1$. For that purpose, we define the linear map
\begin{align}
    \mathcal{G}_t(X) 
    = \sqrt{\rho_{\mathrm{ss}}}\,
      \Phi_t\!\Big((\sqrt{\rho_{\mathrm{ss}}})^{-1} \,
      X \,(\sqrt{\rho_{\mathrm{ss}}})^{-1}\Big)
      \sqrt{\rho_{\mathrm{ss}}}\,,
    \label{eq:EM2}
\end{align}
which represents a weighted transformation of the original map $\Phi_t$ with respect to the steady state $\rho_{\mathrm{ss}}$. This transformation allows us to write,
\begin{align}
\sqrt{\rho_{\mathrm{ss}}}\,\Phi_t(X)\,\sqrt{\rho_{\mathrm{ss}}} 
    = \mathcal{G}_t\!\left(\sqrt{\rho_{\mathrm{ss}}}\,X\,\sqrt{\rho_{\mathrm{ss}}}\right),
    \label{eq:EM3}
\end{align}
which shows the connection between the action of $\Phi_t$ on $X$ with the action of $\mathcal{G}_t$ on the weighted-transformed operator. As a consequence, the dressed distance in Eq.~\eqref{eq:EM1} can be written as,
\begin{align}
 \mathcal{D}_{\rm dd}\big(\Phi_t(\mathcal{O}(0)-\mathcal{O}_{\mathrm{ss}})\big) &= 
 \! |\!| \Phi_t(X) |\!|_{\rho_{\rm{ss}}} =
  |\!|\sqrt{{\rho_{\mathrm{ss}}}} \Phi_t(X)\sqrt{{\rho_{\mathrm{ss}}}}|\!| \nonumber\\
   & = |\!|\mathcal{G}_t\!\left(\sqrt{\rho_{\mathrm{ss}}}\,X\,\sqrt{\rho_{\mathrm{ss}}}\right)|\!|,
    \label{eq:EM4}
\end{align}
where we identify $X=\mathcal{O}(0)-\mathcal{O}_{\rm ss}$. It turns out that the linear map $\mathcal{G}_t$ in Eq.~\eqref{eq:EM2} is a trace-preserving map. This can be shown as follows. For any arbitrary operator $A$, we write 
\begin{align}
\mathrm{Tr}\big[\mathcal{G}_t(A)\big] &=\mathrm{Tr}\Big[\Phi_t\Big((\sqrt{\rho_{\mathrm{ss}}})^{-1}A(\sqrt{\rho_{\mathrm{ss}}})^{-1}\Big)\rho_{\mathrm{ss}}\Big],\nonumber\\
&=\mathrm{Tr}\Big[(\sqrt{\rho_{\mathrm{ss}}})^{-1}A(\sqrt{\rho_{\mathrm{ss}}})^{-1}\Lambda_t(\rho_{\mathrm{ss}})\Big],\nonumber\\
&=\mathrm{Tr}\Big[(\sqrt{\rho_{\mathrm{ss}}})^{-1}A(\sqrt{\rho_{\mathrm{ss}}})^{-1}\rho_{\mathrm{ss}}\Big]=\mathrm{Tr}\big[A\big].
\label{eq:EM5}
\end{align}
Here in the second line $\Lambda_t(\rho_{\mathrm{ss}})$ represents the action of the forward map $e^{\mathcal{L}t}$ on the steady-state density matrix $\rho_{\mathrm{ss}}$.  As $\mathcal{G}_t$ turns out to be a trace-preserving map, overall it becomes a CPTP map and thus inherits the contractive property of which is crucial in establishing the monotonicity of the distance measure. 

We next consider the operator $\mathcal{O}$ to be evolving under the map $\mathcal{G}_t$. Hence we define, $\mathcal{O}_{\mathcal{G}}(t)=\mathcal{G}_t(\mathcal{O})$ and write
$\rm{Tr}[\mathcal{O}_{\mathcal{G}}(t)-\mathcal{O}_{\mathcal{G}}(\infty)] = \rm{Tr}[ \mathcal{G}_t(\mathcal{O}(0))-\mathcal{G}_{\infty}(\mathcal{O}(0))]=\rm{Tr}[\mathcal{O}(0)-\mathcal{O}(0)]=0$, at any time $t$. This shows that trace of $\mathcal{O}_{\mathcal{G}}(t)-\mathcal{O}_{\mathcal{G}}(\infty)$ remains invariant under the map $\mathcal{G}_t$. Moreover, since, $\mathcal{O}_{\mathcal{G}}(t)-\mathcal{O}_{\mathcal{G}}(\infty)$ is also self-adjoint, we can write $\mathcal{O}_{\mathcal{G}}(t)-\mathcal{O}_{\mathcal{G}}(\infty)=Q_t-S_t$, where $Q_t$ and $S_t$ are positive semi-definite matrices with mutually orthogonal support. Let us write the eigen decomposition of $Q_t$ as,
$
Q_t=\sum_i q_{t}^{\,i}\,\ket{q_{t}^{\,i}}\!\bra{q_{t}^{\,i}},
$
and accordingly we define the projector onto the support of $Q_t$ by
$
\mathbb{P}_{t}:=\sum_{i:\;q_{t}^{\,i}\neq 0}\ket{q_{t}^{\,i}}\!\bra{q_{t}^{\,i}}.
$
Now, for any two times \(t_2>t_1\), \(\mathbb{P}_{t_2},\mathbb{P}_{t_1}\) are the projector onto the supports of \(Q_{t_2}\) and \(Q_{t_1}\), respectively. We then construct the standard trace distance measure $D_{\rm tr}$ but involving the operator $\mathcal{O}$ evolving under the map $\mathcal{G}_t$. We therefore write, 
\begin{align}
\mathcal{D}_{\mathrm {tr}}\big(\mathcal{O}_{\mathcal{G}}(t_1),\mathcal{O}_{\mathcal{G}}(\infty)\big) 
&= |\!| \big(\mathcal{O}_{\mathcal{G}}(t_1)-\mathcal{O}_{\mathcal{G}}(\infty)\big)|\!|=\mathrm{Tr}|\mathcal{O}_{\mathcal{G}}(t_1)-\mathcal{O}_{\mathcal{G}}(\infty)|\nonumber\\&= \mathrm{Tr}\,|Q_{t_1}-S_{t_1}| =  \mathrm{Tr}\,\big[Q_{t_1}+S_{t_1}\big] \nonumber\\
&= 2 \, \mathrm{Tr}\,\big[Q_{t_1}\big] 
=2 \,\mathrm{Tr}\big[\mathcal{G}_{t_2} (Q_{t_1})\big]\nonumber\\
&\geq 2 \, \mathrm{Tr}\big[\mathbb{P}_{t_2}\mathcal{G}_{t_2} (Q_{t_1})\big].
\label{eq:EM6}
\end{align}
The first inequality in Eq.~\eqref{eq:EM6} arises because, under the action of the non-unitary map $\mathcal{G}_{t_2}$, the evolution of $Q_{t_1}$, given by $\mathcal{G}_{t_2}(Q_{t_1}) = \widetilde{Q}_{t_2}$, does not, in general, preserve the mutual orthogonal supports with $S_{t_2}$, but whereas $Q_{t_2}$ does. Consequently, $\widetilde{Q}_{t_2}$ possesses a spectral decomposition distinct from that of $Q_{t_2}$, namely,
$
\widetilde{Q}_{t_2} = \sum_{i:\,\widetilde{q}_{t_2}^{\,i}\neq 0} 
\widetilde{q}_{t_2}^{\,i}\,
\ket{\widetilde{q}_{t_2}^{\,i}}\!\bra{\widetilde{q}_{t_2}^{\,i}}.
$
Furthermore, noting that the, $\bra{\widetilde{q}_{t_2}^{\,i}}\mathbb{P}_{t_2}\ket{\widetilde{q}_{t_2}^{\,i}}$ lies between $[0,1]$ i.e., it satisfies this inequality, $0\leq\bra{\widetilde{q}_{t_2}^{\,i}} \mathbb{P}_{t_2}\ket{\widetilde{q}_{t_2}^{\,i}}\leq1$. As a result it justifies the inequality in Eq.~\eqref{eq:EM6}. We can further show that,
\begin{align}
  \mathcal{D}_{\mathrm {tr}}\big(\mathcal{O}_{\mathcal{G}}(t_1),\mathcal{O}_{\mathcal{G}}(\infty)\big) &\geq 2 \, \mathrm{Tr}\big[\mathbb{P}_{t_2}\mathcal{G}_{t_2} (Q_{t_1})\big]\nonumber\\   &\geq 2\,\mathrm{Tr}\big[\mathbb{P}_{t_2}\mathcal{G}_{t_2} (Q_{t_1}-S_{t_1})\big]\nonumber\\
&=2\,\mathrm{Tr}\big[\mathbb{P}_{t_2}\big(\mathcal{O}_{\mathcal{G}}(t_2)-\mathcal{O}_{\mathcal{G}}(\infty)\big)\big]\nonumber\\
&=2\, \mathrm{Tr}\big[\mathbb{P}_{t_{2}}\big(Q_{t_2}-S_{t_2}\big)\big]\nonumber\\
&=2\,\mathrm{Tr}\big[Q_{t_2}\big]
=\mathcal{D}_{\mathrm{tr}}\big(\mathcal{O}_{\mathcal{G}}(t_2),\mathcal{O}_{\mathcal{G}}(\infty)\big).
\label{eq:EM6_1}
\end{align}
Under the action of the map $\mathcal{G}_{t_2}$, the positive semi-definite matrix $S_{t_1}$ evolves to $\widetilde{S}_{t_2}$. 
As a result, we have $\mathbb{P}_{t_2}\widetilde{S}_{t_2} \neq 0$, whereas $\mathbb{P}_{t_2} S_{t_2} = 0$. The second inequality in Eq.~\eqref{eq:EM6_1} arises because we subtract a positive quantity, $\mathrm{Tr}\big[\mathbb{P}_{t_2}\mathcal{G}_{t_2}(S_{t_1})\big]$, from $\mathrm{Tr}\big[\mathbb{P}_{t_2}\mathcal{G}_{t_2}(Q_{t_1})\big]$. The term $\mathrm{Tr}\big[\mathbb{P}_{t_2}\mathcal{G}_{t_2}(S_{t_1})\big]$ is positive since $S_{t_1}$ is a positive semi-definite operator that evolves under a completely positive and trace-preserving (CPTP) map $\mathcal{G}_{t_2}$. Finally we receive the result that the standard trace distance between two operators decreases monotonically under the action of the map~$\mathcal{G}_t$, i.e., 
\begin{align}
\mathcal{D}_{\mathrm{tr}}(\mathcal{O}_{\mathcal{G}}(t_1),\mathcal{O}_{\mathcal{G}}{(\infty)}) \geq \mathcal{D}_{\mathrm{tr}}(\mathcal{O}_{\mathcal{G}}(t_2),\mathcal{O}_{{\mathcal{G}}}(\infty))  ~ \forall \, t_2 \geq t_1.
\label{eq:EM7}
\end{align}
Therefore, following Eq.~\eqref{eq:EM4} and Eq.~\eqref{eq:EM7},
we can write 
\begin{align}
  |\!| \Phi_{t_2}(X) |\!|_{\rho_{\rm{ss}}}  &=|\!|\mathcal{G}_{t_2}(\sqrt{\rho_{\rm{ss}}}X\sqrt{\rho_{\rm{ss}}})|\!|\nonumber\\&\leq |\!|\mathcal{G}_{t_1}(\sqrt{\rho_{\rm{ss}}}X\sqrt{\rho_{\rm{ss}}})|\!|=|\!| \Phi_{t_1}(X) |\!|_{\rho_{\rm{ss}}},
  \label{eq:EM8}
\end{align}
where, $|\!| \,.\, |\!|=\mathrm{Tr}\,| \,.\, |=\mathrm{Tr}\sqrt{(\,.\,)^{\dagger}(\,.\,)}$, is the standard trace distance. From which it follows,
\begin{align}
    |\!| \Phi_{t_2}(X) |\!|_{\rho_{\rm{ss}}}&\leq |\!| \Phi_{t_1}(X) |\!|_{\rho_{\rm{ss}}} \quad  \nonumber\\
    \mathcal{D}_{\mathrm{dd}}(\mathcal{O}(t_2),\mathcal{O}_{\rm{ss}}) &\leq \mathcal{D}_{\rm dd}(\mathcal{O}(t_1),\mathcal{O}_{\rm{ss}}) ~~~\forall ~~~ t_2 \geq t_1.
    \label{eq:EM9}
\end{align}
Therefore, we conclude that the dressed distance under the non-trace preserving (nTP) map $\Phi_t= e^{\mathcal{L}^{\dagger}t}$ decays monotonically.

\section*{B: Accelerated relaxation for operators with the same operator steady state}
In this section, we show that for operators with the same steady state before and after unitary transformation, only accelerated relaxation is possible, and \textit{no} Mpemba-like effect can occur.  If the unitary operator $U$ that is applied to the initial operator $\mathcal{O}(0)$ is chosen such a way that it commutes with the steady-state density matrix $\rho_{\rm ss}$, i.e., $[U, \rho_{\rm ss}]=0$, then it is easy to see that $\widetilde{\mathcal{O}}_{\rm ss}= \mathrm{Tr}\!\left[\rho_{\mathrm{ss}}\, U^{\dagger}\mathcal{O}(0)U\right] \mathbb{I}= \mathcal{O}_{\rm ss}$. It can be further shown that the initial dressed distance before and after the unitary is same, i.e., $\mathcal{D}_\mathrm{dd}(\mathcal{O}(0),\mathcal{O}_{\rm{ss}})=\mathcal{D}_\mathrm{dd}(\widetilde{\mathcal{O}}(0),\mathcal{O}_{\rm{ss}})$. This is what we show below. 

Before unitary, the initial dressed trace distance is,
\begin{equation}
\mathcal{D}_\mathrm{dd}(\mathcal{O}(0),\mathcal{O}_{\rm{ss}})=\rm{Tr}\Big|\sqrt{\rho_{\rm{ss}}}\big(\mathcal{O}(0)-\mathcal{O}_{\rm{ss}}\big)\sqrt{\rho_{\rm{ss}}}\Big|.
\end{equation}
After unitary, the trace distance becomes
\begin{align}
\mathcal{D}_\mathrm{dd}(\widetilde{\mathcal{O}}(0),\mathcal{O}_{\rm{ss}})=\mathrm{Tr}\Big|\sqrt{\rho_{\mathrm{ss}}}\Big(U^{\dagger}\mathcal{O}(0)U-\mathcal{O}_{\rm{ss}}\Big)\sqrt{\rho_{\mathrm{ss}}}\Big|.
 \label{eq:EM10}
\end{align}
Now, if $[U,\rho_{\mathrm{ss}}]=0$, from Eq.~\eqref{eq:EM10} we obtain,
\begin{equation}
\mathcal{D}_\mathrm{dd}(\widetilde{\mathcal{O}}(0),\mathcal{O}_{\mathrm{ss}})=\mathrm{Tr}\Big|U^{\dagger} \sqrt{\rho_{\mathrm{ss}}}\,\Big(\mathcal{O}(0)-{\mathcal{O}}_{\mathrm{ss}}\Big)\sqrt{\rho_{\mathrm{ss}}}\,U\Big|=\mathcal{D}_\mathrm{dd}({\mathcal{O}}(0),\mathcal{O}_{\rm{ss}}).
\label{eq:EM11}
\end{equation}
Since the unitary commutes with the steady state, applying the unitary transformation to the operator is equivalent to applying it to the entire expression in Eq.~\eqref{eq:EM11}. Furthermore, because the trace is invariant under unitary transformation, the dressed distance remains same before and after the unitary transformation. Consequently, the corresponding operator before ($\mathcal{O}(0)$) and after ($\widetilde{\mathcal{O}}(0)$) the unitary transformation start from the same point and converge to the same steady state. As a consequence, only accelerated relaxation remains as a possibility, whereas Mpemba like effect can not occur in such a scenario. A  special case would be a unital map corresponding to the Liouvillian $\mathcal{L}$ where the steady state $\rho_{\rm ss}$ is proportional to  identity. More precisely, $\rho_{\mathrm{ss}}=\mathbb{I}/d$, and as a result, the condition $[U,\rho_{\mathrm{ss}}]=0$ is trivially satisfied. For a given operator, some specific choice of $U$ could then lead to accelerated relaxation.

\section*{C: Absence of Mpemba-like effect for operators in single qubit system}

In this section, we show that for an arbitrary operator in the single-qubit Hilbert space, there can not be any Mpemba-like effect. Any general operator in the single-qubit Hilbert space can be written as 
$\mathcal{O} = b_0 \, \mathbb{I} + \vec{b} \cdot \vec{\sigma}$, 
where $\vec{b} \in \mathbb{R}^3$ is a real three-dimensional vector and 
$\vec{\sigma} = (\sigma_x, \sigma_y, \sigma_z)$ is the vector of Pauli matrices. 
In the case of dissipators $\sigma_{+}$ and $\sigma_{-}$, 
the right eigenmatrices $r_1$ and $r_2$ corresponding to the slowest decaying modes 
$\lambda_1$ and $\lambda_2=\lambda_1^{*}$ 
are purely off-diagonal. 
To achieve an acceleration in the operator dynamics, 
one can perform a unitary transformation on $\mathcal{O}$ such that it becomes diagonal. 
The transformed operator then takes the form 
$\widetilde{\mathcal{O}} = \begin{pmatrix} b_0 + b & 0 \\ 0 & b_0 - b \end{pmatrix}$
where $b = \sqrt{b_x^2 + b_y^2 + b_z^2}$. As a result, the operator will quickly relax to its respective steady state. To investigate how its initial dressed distance relative to the steady state changes after the unitary operation, we compute the dressed distance for the two cases. In our setup, the steady state is a detailed-balance thermal state, given by $\rho_{\mathrm{ss}} = e^{-\beta H}/\mathcal{Z}$, where $H=\frac{\omega_0}{2} \sigma_z$, $\beta = 1/(k_B T)$ and the partition function $\mathcal{Z} = 2\cosh{(\beta \omega_0 / 2)}$. 

The dressed distance, following Eq.~\eqref{dressed_norm} of the main text, can be written as  
$
\mathcal{D}_\mathrm{dd}(\mathcal{O}_0, \mathcal{O}_{\mathrm{ss}}) = \sum_{i=1}^{2} |\lambda_i'|
= 2\sqrt{b_x^2 + b_y^2 + (2b_z/\mathcal{Z})^2}/\mathcal{Z},
$
where $\lambda_i'$ are the eigenvalues of the operator 
$\rho_{\mathrm{ss}}^{1/2} (\mathcal{O}_0 - \mathcal{O}_{\mathrm{ss}}) \rho_{\mathrm{ss}}^{1/2}$.  
After the unitary transformation, the distance between the transformed operator and its corresponding steady state becomes  
$
\mathcal{D}_\mathrm{dd}(\widetilde{\mathcal{O}}_0, \widetilde{\mathcal{O}}_{\mathrm{ss}}) 
= \sum_{i=1}^{2} |\widetilde{\lambda}_i'|
= 4b/\mathcal{Z}^2,
$
which is always smaller than the dressed distance before the unitary transformation, $\mathcal{D}_\mathrm{dd}(\mathcal{O}_0, \mathcal{O}_{\mathrm{ss}})$.
Here, the unitary $U$ removes the overlap with the slowest decaying modes, and $\widetilde{\lambda}_i'$ denote the eigenvalues of 
$\rho_{\mathrm{ss}}^{1/2} (\widetilde{\mathcal{O}}_0 - \widetilde{\mathcal{O}}_{\mathrm{ss}}) \rho_{\mathrm{ss}}^{1/2}$.  
As a result, no Mpemba-like effect can be observed in the single-qubit case with the chosen dissipators. Note that, instead of making zero overlap with the slowest decay modes, one can reduce the overlap via choosing an appropriate unitary transformation. However, in these cases also, the dressed distance post-unitary is always smaller than before unitary. Hence, the conclusion of no Mpemba-like effect for single qubit setup still holds. 
\vspace{0.1cm}

\section*{D: Details about unitary protocols to observe accelerated relaxation or Mpemba-like effect}
In this section, we provide the necessary details about choosing the unitary transformation such that accelerated relaxation or Mpemba-like effect can be observed for operator relaxation. To get an acceleration or speed up the relaxation process, we have to bypass or reduce the overlap of the operator with the slowest decay mode by choosing a suitable unitary transformation. 
In our first example in the main text, i.e., for the single-qubit case, we choose a rotation matrix as our unitary operator $U=e^{i\theta\sigma_i}$ where $\sigma_i=\{\sigma_x,\sigma_y\}$ when the corresponding chosen operators are $\mathcal{O}_i=\{\sigma_y,\sigma_x\}$.

In the second example, i.e., for the qutrit case, we were interested in the Mpemba-like effect for the Hamiltonian operator and in this case we choose a swap operator as our unitary operator having the form 
$U = \{(0,1,0),(1,0,0),(0,0,1)\}$ to reduce the overlap with the first non-zero slowest decay mode for the Hamiltonian $H$ which is in this case $r_3$.

For the third example, i.e., the double quantum dot (DQD) model, we choose the current operator.In this case, to observe the Mpemba-like effect, the unitary operator is constructed from the eigenvectors of the current operator, which diagonalize it.

To obtain accelerated relaxation for the current operator with the same steady state, we choose our unitary such that it commutes with $\rho_{\mathrm{ss}}$. Now for the DQD setup, the steady-state density matrix is not diagonal. In fact, it has non-zero coherence elements  
$[\rho_{\mathrm{ss}}]_{23}$ and $[\rho_{\mathrm{ss}}]_{32}$,
in the steady-state, which are proportional to the inter-dot coupling strength $g$. Therefore the choice of unitary $U$ can be made by constructing $U = W V W^{\dagger}$ 
where  $W$ is the unitary constructed from the eigenvectors of $\rho_{\mathrm{ss}}$, and $V$ is a diagonal unitary having the form 
$V = \mathrm{diag}(e^{i\theta_1}, e^{i\theta_2}, e^{i\theta_3}, e^{i\theta_4})$. Such a $U$
will commute with the steady state $\rho_{\rm ss}$.
In the simulation, the corresponding unitary parameters for which the overlap with the slowest decay mode is reduced are $\theta_1 = \theta_4=0$, $\theta_2 = \pi/8$, and $\theta_3 = 2.5583$.

\begin{figure}
    \centering
\includegraphics[trim=0cm 0cm 0cm 0cm, clip=true,width=7.5cm]{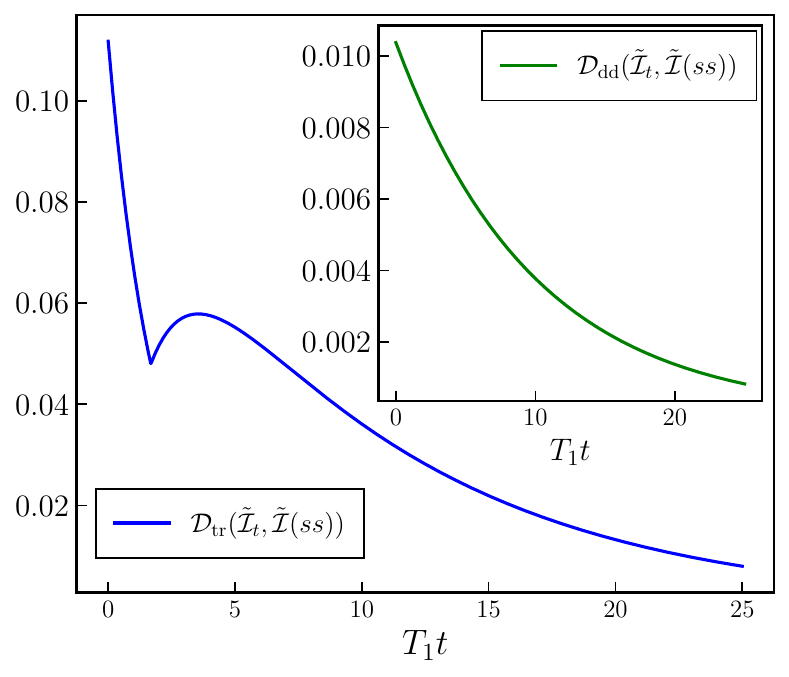}
\caption{Non-monotonic decay of the trace distance $\mathcal{D}_{\mathrm {tr}}$ for unitarily transformed energy current operator $\tilde{I}$ for the double-quantum dot (DQD) setup, as introduced in example (3) of the main text. The inset shows the plot for dressed distance $\mathcal{D}_{\mathrm {dd}}$ involving the same operator, which decays monotonically.  Simulations are done with parameter values    $\epsilon_{\rm{d1}}=2$, $\epsilon_{\rm{d2}}=1$, $g=0.05$, $T_1=1$, $\mu_1=0$, $T_2=0.1$, $\mu_2=0$, $\gamma_1=0.1$, $\gamma_2=0.5$.}
\label{non-monotonicity}
\end{figure}

\section*{E: Non-monotonic decay of trace distance for operators under CPTP map.--}

In this section, we show that the trace distance measure, which is typically used to detect the Mpemba effect for quantum state relaxation, is not a good measure for detecting the Mpemba-like effect for operators. The evolution of the density matrix is governed by the map $e^{\mathcal{L} t}$ where $\mathcal{L}$ is the Liouvillian, is trace preserving. As a result, the trace distance between any two density matrices is a monotonically decaying function of time when evolving under $\mathcal{L}$. However, as the operator's evolution is governed by the map $e^{\mathcal{L}^{\dagger}t}$, which is, in general, a non-trace preserving map, the trace distance of an operator to the one to which it converges in the long time limit, is not a monotonically decaying function with time. As a consequence, the trace distance is not a good measure to detect Mpemba-like effects in operators.
In contrast, the steady-state dressed distance, as introduced in Eq.~\eqref{dressed_norm} is shown to always decay monotonically under the map $e^{\mathcal{L}^{\dagger}t}$. The proof holds universally for arbitrary Hermitian operator $\mathcal{O}$ and arbitrary dissipators. 

We exemplify this point for the double-quantum dot (DQD) setup, as discussed in Example 3 of the main text. We calculate both the trace distance, defined as 
\begin{equation}
\mathcal{D}_{\mathrm {tr}}\big(\mathcal{O}(t),\mathcal{O}(\infty)\big) 
= \mathrm{Tr}\,\Big| \mathcal{O}(t)- \mathcal{O}(\infty)\Big|
\end{equation}
and the dressed distance, defined in Eq.~\eqref{dressed_norm}, for the energy current operator $I=ig\Big(\sigma^{(1)}_- \otimes\sigma^{(2)}_+-\sigma^{(1)}_+ \otimes \sigma^{(2)}_-\Big)$. We perform a unitary transformation to the operator $I$ to make it diagonal $\tilde{I}=\mathrm{diag}(-g,g,0,0)$ and calculate the distance measures. Recall that $g$ is the inter dot coupling strength. 
We found that, the trace distance measure  decay non-monotonically, as shown in Fig.~\ref{non-monotonicity}, while for the dressed distance the decay is monotonic (as shown in the inset of Fig.~\ref{non-monotonicity}).

\end{document}